\documentclass[12pt]{article}
\pdfoutput=1

\usepackage{amsmath}
\usepackage{amsfonts}
\usepackage{amssymb}
\usepackage{dsfont}

\usepackage[
      colorlinks=true,
      linkcolor=blue,
      urlcolor=blue,
      filecolor=black,
      citecolor=red,
      pdfstartview=FitV,
      pdftitle={},
      pdfauthor={Kevin Chen, Michael Gutperle, Christoph Uhlemann},
      bookmarksopen=true
      ]{hyperref}

\def\cS{{\cal S}}

\def\tr{{\rm tr}}

\def\ZZ{{\mathds{Z}}}

\usepackage{color}
\usepackage{graphicx}

\usepackage{sectsty}
\sectionfont{\large}

\renewcommand{\thefootnote}{\fnsymbol{footnote}}
\renewcommand{\thanks}[1]{\footnote{#1}}
\newcommand{\starttext}{
\setcounter{footnote}{0}
\renewcommand{\thefootnote}{\arabic{footnote}}}
\newcommand{\bea}{\begin{eqnarray}}
\newcommand{\eea}{\end{eqnarray}}
\newcommand{\be}{\begin{equation}}
\newcommand{\ee}{\end{equation}}

\numberwithin{equation}{section}

\newcommand{\N}[1]{\ensuremath{\mathcal N=#1}}

\long\def\symbolfootnote[#1]#2{\begingroup%
\def\thefootnote{\fnsymbol{footnote}}\footnote[#1]{#2}\endgroup}

\begin{document}
\setlength{\baselineskip}{16pt}

\starttext
\setcounter{footnote}{0}

\begin{flushright}
\today
\end{flushright}

\bigskip

\begin{center}

{\Large \bf Spin 2 operators in holographic 4d $\N{2}$ SCFTs}

\vskip 0.4in

{\large  Kevin Chen, Michael Gutperle, Christoph F.~Uhlemann}

\vskip 0.2in

{ \sl Mani L. Bhaumik Institute for Theoretical Physics} \\
{\sl Department of Physics and Astronomy }\\
{\sl University of California, Los Angeles, CA 90095, USA} 

\bigskip

\end{center}
 
\begin{abstract}
\setlength{\baselineskip}{16pt}
A broad class of holographic duals for 4d $\N{2}$ SCFTs is based on the general half-BPS $AdS_5$ solutions to M-theory constructed by Lin, Lunin and Maldacena, and their Kaluza-Klein reductions to Type IIA. We derive the equations governing spin 2 fluctuations around these solutions. The resulting partial differential equations admit families of universal solutions which are constructed entirely out of the generic data characterizing the background. This implies the existence of families of spin 2 operators in the dual SCFTs, in short superconformal multiplets which we identify.
\end{abstract}

\setcounter{equation}{0}
\setcounter{footnote}{0}

\newpage

\section{Introduction and summary}

Four-dimensional \N{2} superconformal field theories (SCFTs) have been studied extensively.
They are less constrained than \N{4} theories while retaining sufficient structure for powerful field theory methods to be applicable.
Their holographic description is based on $AdS_5$ solutions preserving sixteen supersymmetries. The general form of such solutions in M-theory has been constructed in \cite{Lin:2004nb}, and contains $S^2$ and $S^1$ spaces in addition to $AdS_5$ to realize the bosonic part of the field theory supersymmetry. 
Holographic duals for the celebrated 4d class $\cS$ theories \cite{Gaiotto:2009we} have been identified within this class of solutions in \cite{Gaiotto:2009gz}.
Further aspects have been studied in \cite{Donos:2010va,Petropoulos:2013vya,Petropoulos:2014rva}.
A generic solution is specified entirely by a solution $D$ to a 3d Toda equation in the remaining coordinates $x_1$, $x_2$, $y$,
\begin{align}\label{eq:toda}
 \left(\partial_{x_1}^2+\partial_{x_2}^2\right)D+\partial_y^2 e^D&=0~.
\end{align}
The data specifying the field theory is encoded in the boundary conditions imposed on $D$ and the equation may be supplemented by source terms encoding additional data, which will be discussed in more detail below.
Upon assuming a $U(1)$ isometry in the $x_1,x_2$ plane, the solutions may be reduced to Type IIA. 
After a change of variables the Type IIA solutions are then characterized by a solution $V$ to the three-dimensional cylindrical Laplace equation in $\sigma$, $\eta$,
\begin{align}\label{eq:laplace}
 \frac{1}{\sigma}\partial_\sigma\sigma\partial_\sigma V + \partial_\eta^2 V &= 0~.
\end{align}
Field theories with holographic duals within this class include many 4d quiver SCFTs engineered by brane constructions involving D4, NS5 and D6 branes, and have been studied e.g.~in \cite{ReidEdwards:2010qs,Aharony:2012tz,Lozano:2016kum,Nunez:2019gbg}.

In this note we identify a universal feature of the 4d \N{2} SCFTs with holographic duals of the aforementioned types. 
Namely, the spectrum of gauge-invariant operators includes, in addition to the conserved energy-momentum tensor, families of spin 2 operators with scaling dimensions
\begin{align}\label{eq:Delta}
 \Delta&=4+2\ell+n~, & \ell, n&\in \ZZ~, &\ell, n &\geq 0~,
\end{align}
with $\ell$ and $n$ specifying the $SU(2)$ and $U(1)$ $R$-charges, respectively.
These operators are $Q^2\bar Q^2$ descendants in short \N{2} superconformal multiplets which we identify explicitly in sec.~\ref{sec:multiplets}.
In the gravitational description they correspond to universal spin 2 fluctuations that are constructed entirely out of the background data (the function $D$ in M-theory and $V$ in Type IIA), and therefore exist in generic background solutions of the form constructed in~\cite{Lin:2004nb}.

Constructing linearized fluctuations around the supergravity backgrounds of \cite{Lin:2004nb} in general is a non-trivial problem. 
The linearized supergravity field equations reduce to coupled second-order partial differential equations in $(x_1,x_2,y)$ in M-theory and in $(\sigma,\eta)$ in Type IIA.
For spin 2 fluctuations the problem simplifies, as they arise solely from metric perturbations and are determined by a decoupled equation depending only on the geometry of the background solution \cite{Bachas:2011xa}.
This has been exploited in other contexts e.g.\ in \cite{Klebanov:2009kp,Richard:2014qsa,Passias:2016fkm,Schmude:2016bqp,Pang:2017omp}.
For the solutions of \cite{Lin:2004nb}, this still leaves a non-trivial second-order PDE to be solved. 
A massless spin 2 fluctuation is expected for all solutions, as holographic dual to the SCFT energy-momentum tensor. But beyond that the spectrum may depend on the specific solution at hand, reflecting more detailed features of the dual SCFT.
Given the large class of field theories that are expected to be described by the solutions of \cite{Lin:2004nb}, the existence of further universal fluctuations therefore is an interesting result.
We note that the operators (\ref{eq:Delta}) are reminiscent of universal spin 2 operators in 5d SCFTs \cite{Passias:2018swc,Gutperle:2018wuk}, corresponding to universal fluctuations around the solutions of \cite{Brandhuber:1999np,DHoker:2016ujz,DHoker:2017mds,DHoker:2017zwj}. It would be interesting to better understand this relation, e.g.\ through circle compactification of the 5d SCFTs.
More generally, one may wonder whether similar universal fluctuations can be identified e.g.\ for the more general M-theory solutions of \cite{Gauntlett:2004zh}, the M-theory holographic duals for 4d \N{1} SCFTs of \cite{Bah:2012dg}, or the massive Type IIA solutions of \cite{Apruzzi:2015zna}.

The paper is organized as follows. In sec.~\ref{sec:M-theory} we review the M-theory solutions of \cite{Lin:2004nb} and construct the spin 2 fluctuations. In sec.~\ref{sec:IIA} the analogous construction is presented in Type IIA. 
In sec.~\ref{sec:multiplets} we identify the superconformal multiplets containing the spin 2 fluctuations and comment briefly on the field theory realization of the dual operators.

\section{Spin 2 fluctuations on \texorpdfstring{$AdS_5$}{AdS5} in M-theory}\label{sec:M-theory}

In section \ref{sec:M-theory-review} we  review the M-theory solutions  dual to 4d  $\N{2}$ SCFTs constructed in  \cite{Lin:2004nb,Gaiotto:2009gz}.   The  transverse traceless spin 2 fluctuations in the $AdS_5$ space are constructed  in sec.~\ref{sec:spin-2-M-theory}. Regularity  of the fluctuations is discussed in sec.~\ref{sec:norm-M} and \ref{sec:M2-bound}.

\subsection{LLM solutions in M-theory}\label{sec:M-theory-review}

The geometry in the M-theory solutions of \cite{Lin:2004nb} contains $AdS_5$, an $S^2$ and an $S^1$ which  is parametrized by $\beta$. 
In a form which will be convenient later, the metric reads
\begin{align}\label{eq:metric-M}
 ds^2&=4f_1 ds^2_{AdS_5}+\hat g_{ab}dz^a dz^b~,
&
 \hat g_{ab}dz^a dz^b&=f_2ds^2_{S^2}+f_4 dy^2+f_6ds_3^2~,
 \nonumber\\
 && ds_3^2&=dx_i dx^i+\frac{f_3}{f_6}\left(d\beta+A_i dx^i\right)^2~,
\end{align}
where $i=1,2$ and the coordinates on the internal space, namely those comprising $S^2$, $y$ and $(x_1,x_2,\beta)$, are collectively denoted by $z^a$. The metric functions are 
\begin{align}\label{metfacm}
 f_1&=e^{2\tilde\lambda}~, & f_2&=y^2 e^{-4\tilde\lambda}~, &
 f_3&=4e^{2\tilde\lambda}(1-y^2e^{-6\tilde\lambda})~, \nonumber\\
 f_4&=\frac{e^{-4\tilde\lambda}}{1-y^2e^{-6\tilde\lambda}}~, & 
 f_6&=f_4e^D~,
 &
 A_i&=\frac{1}{2}\epsilon_{ij}\partial_j D~.
\end{align}
The function $\tilde\lambda$ is defined in terms of a function $D(x_1,x_2,y)$ by
\begin{align}\label{lamfac}
 e^{-6\tilde \lambda}&=-\frac{\partial_y D}{y(1-y\partial_y D)}~.
\end{align}
The expression for the four-form $F_4$ will not be needed here and can be found in \cite{Lin:2004nb}.
It was shown in  \cite{Lin:2004nb}  that this supergravity background satisfies the equation of motion of eleven dimensional supergravity and preserves sixteen real supersymmetries if the function $D$ satisfies the 3d Toda equation in the $x_1,x_2$ and $y$ coordinates,
\begin{align}\label{eq:toda-LLM}
 (\partial_1^2+\partial_2^2)D+\partial_y^2 e^D&=0~.
\end{align}

\subsubsection{Boundary conditions}
Holographic duals for large classes of 4d \N{2} SCFT based on the solutions of \cite{Lin:2004nb} were constructed  in \cite{Gaiotto:2009gz}.
Boundary conditions are imposed at $y=0$ such that the $S^2$ collapses, which is realized by
\begin{align}
 \partial_yD\vert_{y=0}&=0~, &  D\vert_{y=0}= &\text{ finite}~. \label{eq:bc}
\end{align}
In addition, the $S^1$ collapses at a location $y=y_c$. This produces a 4-cycle from $S^2\times S^1$ warped over the interval $I_y=\lbrace y\vert 0\leq y \leq y_c\rbrace$. The required behavior for $D$ at $y=y_c$ is
\begin{align}\label{eq:bc-yc}
 e^D\vert_{y\sim y_c}\sim y-y_c~.
\end{align}
Explicit M5-brane sources may be added as line sources localized at $x=x^{(i)}$ in the $(x_1,x_2)$ plane in the Toda equation (\ref{eq:toda-LLM}), such that it becomes
\begin{align}\label{eq:toda-LLM-sources}
 (\partial_1^2+\partial_2^2)D+\partial_y^2 e^D&=-2\pi\sum_{i}\delta^{(2)}(x-x^{(i)})\theta(2N_5^{(i)}-y)~.
\end{align}
If explicit sources are added the solution to the Toda equation is singular, but the full metric remains regular \cite{Gaiotto:2009gz}. The behavior of $D$ near the line charges is
\begin{align}\label{eq:near-source}
 e^D\vert_{x\sim x^{(i)}}&\sim\frac{1}{|x-x^{(i)}|}~.
\end{align}

\subsection{Spin 2 fluctuations}\label{sec:spin-2-M-theory}

Consider a transverse traceless symmetric fluctuation $h$ along the $AdS_5$ part of the eleven dimensional metric
\begin{align}\label{spin-2-M}
ds^2 &= 4f_1\left(ds_{AdS_5}^2 + h_{\mu\nu}dx^\mu dx^\nu\right)+\hat g_{ab}dz^a dz^b~,
\end{align}
where $h$ decomposes into a transverse traceless fluctuation on unit radius $AdS_5$, $h_{\mu\nu}^{[tt]}$, which satisfies the equation of motion for a (massive) spin 2 field, and a mode on the internal space,
\begin{align}\label{eq:h-eom}
 h_{\mu\nu}(x,z)&=h^{[tt]}_{\mu\nu}(x) \psi(z)~,&
 \square^{(2)}_{AdS_5}h^{[tt]}_{\mu\nu}&=(M^2-2) h^{[tt]}_{\mu\nu}~.
\end{align}
$\square^{(2)}$ denotes the Laplace operator acting on rank 2 tensors.
As shown in \cite{Bachas:2011xa}, the linearized Einstein equations then reduce to an eleven-dimensional scalar Laplace equation
\begin{align}\label{eq:h-gen-eom}
 \frac{1}{\sqrt{-g}}\partial_M\sqrt{-g} g^{MN}\partial_N h_{\mu\nu}&=0~.
\end{align}
For the background metric (\ref{eq:metric-M}), with $h$ as in (\ref{eq:h-eom}), this equation reduces to
\begin{align}\label{eq:spin-2-eq-M-0}
\frac{4}{f_1^{3/2}\sqrt{\hat g}} \partial_a \left[f_1^{5/2}\sqrt{\hat g} \hat g^{ab}  \partial_b\right] \psi&=-M^2 \psi~.
\end{align}
The inverse metric on the internal part of the geometry is given by
\begin{align}
 \hat g^{-1}&=\frac{1}{f_2}g^{-1}_{S^2}+\frac{1}{f_4}\partial_y\otimes \partial_y
 +\frac{1}{f_6}g^{-1}_3~,
 \nonumber\\
 g^{-1}_3&=\begin{pmatrix} \partial_{x_1}\\ \partial_{x_2} \\ \partial_\beta\end{pmatrix}^T\otimes
 \begin{pmatrix}
   1 & 0 & -A_1\\
   0 & 1 & -A_2\\
   -A_1 & -A_2 & A_1^2+A_2^2+\frac{f_6}{f_3}
 \end{pmatrix}
 \begin{pmatrix} \partial_{x_1}\\ \partial_{x_2} \\ \partial_\beta\end{pmatrix}~.
\end{align}
The determinant is
\begin{align}
 \sqrt{\hat g}&=f_2f_4^{1/2}f_6 f_3^{1/2}\sqrt{g_{S^2}}~.
\end{align}
Separating the individual parts of the internal space in (\ref{eq:spin-2-eq-M-0}) yields
\begin{align}\label{eq:spin-2-eq-M}
 \frac{4f_1}{f_2}\nabla^2_{S^2}\psi + \frac{4}{f_1^{3/2}\sqrt{\hat g}}\partial_y\left[ \frac{f_1^{5/2}}{f_4}\sqrt{\hat g}\partial_y \right]\psi
 +\frac{4}{f_1^{3/2}\sqrt{\hat g}}\partial_m\left[ \frac{f_1^{5/2}}{f_6}\sqrt{\hat g}g_3^{mn}\partial_n\right]\psi+M^2\psi&=0~,
\end{align}
where $\nabla^2_{S^2}$ denotes the Laplacian on a sphere of unit radius.
The relevant combinations of metric factors are
\begin{align}
 \frac{f_1}{f_2}&=\frac{e^{6\tilde\lambda}}{y^2}~,
 &
 f_1^{3/2}\sqrt{\hat g}&=-2y\partial_y e^D\sqrt{g_{S^2}}~,
 \nonumber\\
 \frac{f_1^{5/2}}{f_4}\sqrt{\hat g}&=2e^D y^2\sqrt{g_{S^2}}~,
 &
 \frac{f_1^{5/2}}{f_6}\sqrt{\hat g}&=2y^2\sqrt{g_{S^2}}~.
\end{align}
Eq.~(\ref{eq:spin-2-eq-M}) thus evaluates to
\begin{align}\label{eq:spin-2-eq-M-2}
 \left[
 \frac{4e^{6\tilde\lambda}}{y^2}\nabla^2_{S^2}
 -\frac{4}{y\partial_y e^D}\partial_y y^2 e^D\partial_y
 -\frac{4y}{\partial_y e^{D}}\partial_m g_3^{mn}\partial_n
 +M^2
 \right]\psi&=0~.
\end{align}
Since $\partial_1 A_1+\partial_2 A_2=0$, 
\begin{align}
 \partial_m g_3^{mn}\partial_n&=\partial_1^2+\partial_2^2
 -2\left(A_1\partial_1+A_2\partial_2\right)\partial_\beta
 +\left(A_1^2+A_2^2+\frac{f_6}{f_3}\right)\partial_\beta^2~.
\end{align}
The function $\psi$ is now expanded in harmonics on $S^2$ and on the $S^1$ parametrized by $\beta$,
\begin{align}\label{eq:psi-exp}
 \psi&=\sum_{\ell m n}\phi_{\ell m n}Y_{\ell m}e^{in\beta}~,  & \nabla^2_{S^2} Y_{\ell m}=-\ell(\ell+1)Y_{\ell m}~,
\end{align}
where $\ell,m,n\in\ZZ$ with $\ell\geq 0$.
$\psi$ has been complexified, with the understanding that the real and imaginary parts provide separate solutions.
With the metric factors (\ref{metfacm}) and the expression for $\tilde \lambda$ given in (\ref{lamfac}), equation (\ref{eq:spin-2-eq-M-2}) becomes
\begin{align}\label{eq:spin-2-eq-M-3}
\Bigg[&
-\frac{4y}{\partial_y e^D}\left(\frac{1}{y^2}\partial_y y^2 e^D\partial_y + \partial_1^2+\partial_2^2-2in\left(A_1\partial_1+A_2\partial_2\right)\right)
\nonumber\\
&
+\frac{4n^2 y}{\partial_y e^D}\left(A_1^2+A_2^2\right)+n^2(y\partial_y D)
+\frac{4\ell(\ell+1)}{y\partial_y D}
+M^2-4\ell(\ell+1)-n^2
\Bigg]\phi_{\ell m n}=0~.
\end{align}
To further simplify this equation we use the following ansatz
\begin{align}
 \phi_{\ell m n}&=y^\ell e^{\frac{n}{2}D}\chi_{\ell m n}~, & M^2&=\mu^2+(2 \ell+n) (2 \ell+n+4)~,
\end{align}
with which eq.~(\ref{eq:spin-2-eq-M-3}) becomes
\begin{align}\label{eq:spin-2-eq-M-4}
\left[
\frac{4y}{e^{(n+1) D}\partial_y D}\left(\frac{1}{y^{2+2\ell}}\partial_y y^{2+2\ell} e^{(n+1)D}\partial_y +\left(\partial_1-i\partial_2\right) e^{n D}\left(\partial_1+i\partial_2\right)\right)-\mu^2
\right]\chi_{\ell m n}&=0~.
\end{align}
This is the general form of the equation satisfied by spin 2 fluctuations.
We now present solutions which are constructed entirely out of the generic background data, i.e.\ the function $D$. Their regularity will be discussed in the next section.
Eq.~(\ref{eq:spin-2-eq-M-4}) is solved by
\begin{align}
\chi_{\ell m n}&=\varphi(x_1,x_2)~, &  \mu^2&=0~,
\end{align}
where $\varphi$ satisfies
\begin{align}
 \left(\partial_1-i\partial_2\right) e^{n D}\left(\partial_1+i\partial_2\right)\varphi&=0~.
\end{align}
This equation is solved when $\varphi$ is holomorphic in $x_1+ix_2$. 
For $n=0$ the complete solution is given by harmonic functions in $x_1$, $x_2$.
In terms of $\phi_{\ell m n}$ and $M^2$, these solutions become
\begin{align}\label{eq:sol}
 \phi^{\rm a}_{\ell m n}&=y^\ell e^{\frac{n}{2}D}\varphi(x_1,x_2)~, & M^2&=-4+(2+2\ell+n)^2~.
\end{align}
The range of $n$ was not constrained in (\ref{eq:psi-exp}); the equation of motion is solved for arbitrary $n\in\ZZ$.
The complex conjugate $\psi^* = Y_{\ell m } e^{-i n \beta} (\phi^{\rm a}_{\ell m n})^* $ solves (\ref{eq:spin-2-eq-M-0}) with the same $M^2$.
In addition, eq.~(\ref{eq:spin-2-eq-M-3}) is invariant under $\ell\rightarrow -\ell-1$, such that there is a second set of solutions
\begin{align}\label{eq:phi-b}
 \phi^{\rm b}_{\ell m n}&=y^{-\ell-1} e^{\frac{n}{2}D}\varphi(x_1,x_2)~, & M^2&=-4+(n-2\ell)^2~.
\end{align}
Finally, there is a solution
\begin{align}\label{eq:phi-c}
 \phi^{\rm c}&=\frac{D}{y}~, & \ell&=n=0~, & M^2&=-4~.
\end{align}
The scaling dimension of the dual operator and the corresponding source for these fluctuations are extracted from $M^2=\Delta(\Delta-4)$, which yields
\begin{align}
\Delta^{\rm a}_{\pm}&=2\pm (2+2\ell+n)~,
&
\Delta^{\rm b}_\pm&=2\pm(n-2\ell)~, & \Delta^{\rm c}_\pm&=2~.
\end{align}
These scaling dimensions are not all compatible with the unitarity bound for spin 2 operators in four dimensions. However, the regularity analysis in the next section will show that those fluctuations which are regular do correspond to operators in unitary representations.

\subsection{Regularity conditions}\label{sec:norm-M}

One can ask whether the metric fluctuations identified in the previous section are finite at generic points of the geometry, including at the boundary $y=0$, where the $S^2$ shrinks and $D$ behaves as in (\ref{eq:bc}). 
The solutions $\phi^{\rm a}$ in (\ref{eq:sol}) are finite at $y=0$ for all $\ell$ and $n$.
The solutions  $\phi^{\rm b}$ of (\ref{eq:phi-b}) and  $\phi^{\rm c}$ of (\ref{eq:phi-c}) are not.\footnote{Unless $D$ is constant at $y=0$, linear combinations of $\phi^{\rm c}$ and $\phi_{000}^{\rm b}$ are not finite at $y=0$ either.} The solutions which are finite are thus $\phi^{\rm a}$.
Finiteness as $y\rightarrow y_c$, with (\ref{eq:bc-yc}), restricts the range of $n$ to $n\geq 0$.
To summarize, the potentially suitable solutions are $\phi^{\rm a}_{\ell m n}$ with $\ell,n\geq 0$, leading to
\begin{align}\label{eq:sol-2}
 \psi_{\ell m n}&= Y_{\ell m} e^{i n \beta} y^\ell e^{\frac{n}{2}D}\varphi(x_1,x_2)~, & \Delta &= 4 + 2\ell + n~, & \ell,n \geq 0~,
\end{align}
and the complex conjugate solutions. Reality of the metric requires real $\psi$, such that the actual solutions for the metric perturbation are given by the real/imaginary parts of $\psi$.
The constraints from the equation of motion on $\varphi$ are different for $n=0$ and $n\neq 0$. 
However, the only holomorphic or harmonic functions which are bounded in the $(x_1,x_2)$ plane are constant. 
Likewise, if $(x_1,x_2)$ parametrize a compact Riemann surface $\varphi$ is expected to be constant for the metric fluctuation to be bounded, since holomorphic and harmonic functions defined everywhere on a compact Riemann surface are constant by the maximum principle.

We now compute the norm of the solutions, starting from the 11d action
\begin{align}
 S&=\frac{1}{2\kappa_{11}^2}\int d^{11}x\sqrt{-g} R + \dots~.
\end{align}
Expanding to second order in a perturbation as set up in (\ref{spin-2-M}) yields
\begin{align}\label{eq:phi-a-rep}
 \delta^2S&=\frac{1}{\kappa_{11}^2}\int d^{11}x h^{\mu\nu}\partial_M\sqrt{-g}g^{MN}\partial_N h_{\mu\nu}
 \nonumber\\
 &=\frac{1}{\kappa_{11}^2}\int d^{11}x\sqrt{-g_{AdS_5}}(4f_1)^{3/2}\sqrt{\hat g} h^{\mu\nu}\left[\square^{(2)}_{AdS_5}+2+\hat\square\right] h_{\mu\nu}~,
\end{align}
where the indices are raised with the metric on unit-radius $AdS_5$, and $\hat\square$ is the differential operator on the left-hand side of (\ref{eq:spin-2-eq-M-0}). Using the expansion (\ref{eq:psi-exp}) for $h_{\mu\nu}$,
\begin{align} \label{eq:M-norm-expand}
h_{\mu\nu} = \sum_{\ell m n} (h^{[tt]}_{\ell m n})_{\mu\nu} \phi_{\ell m n} Y_{\ell m} e^{in \beta}~,
\end{align}
with the spherical harmonics normalized as $\int_{S^2}Y_{\ell m}Y_{\ell^\prime m^\prime}=4\pi \delta_{\ell \ell^\prime}\delta_{m m^\prime}$ and $\beta\in(0,2\pi)$ yields
\begin{align}
 \delta^2S&=\sum_{\ell m n}C_{\ell m n}\int d^5x \sqrt{-g_{AdS_5}}(h_{\ell m n}^{[tt]})^{\mu\nu}\left[\square^{(2)}_{AdS_5}+2-M_{\ell m n}^2\right] (h_{\ell m n}^{[tt]})_{\mu\nu}~,
 \nonumber\\
 C_{\ell m n}&=\frac{8\pi^2}{\kappa_{11}^2}\int d^{3}x(4f_1)^{3/2}\sqrt{\hat g}|\phi_{\ell m n}|^2~.
\end{align}
Explicit evaluation shows
\begin{align}\label{eq:C-M}
 C_{\ell m n}&=-\frac{2^7\pi^2}{\kappa_{11}^2}\int dx_1 dx_2 dy|\phi_{\ell m n}|^2y\partial_ye^D~.
\end{align}
The measure is finite at $y=0$. For the a-type solutions we find
\begin{align}
 C_{\ell m n}^{\rm a}&=-\frac{2^7\pi^2}{\kappa_{11}^2}\int dx_1 dx_2 dy |\varphi|^2 y^{2\ell+1}e^{n D}\partial_ye^D~.
\end{align}
If explicit M5-brane sources are added, as in (\ref{eq:toda-LLM-sources}), the norm of the $n>0$ solutions receives infinite contributions from the regions near the sources due to (\ref{eq:near-source}) unless $\varphi$ vanishes at the sources. 
This may restrict the regular solutions to $n=0$ in this case, although the supergravity approximation breaks down  in the regions near the sources due to large curvatures \cite{Gaiotto:2009gz} and our  conclusions may be subject to corrections.

As an explicit example we discuss the solutions of \cite{Maldacena:2000mw}, describing M5-branes wrapped on a constant curvature Riemann surface of genus $g> 1$. These solutions correspond to \cite{Lin:2004nb}
\begin{align}\label{eq:MN}
 e^D&=\frac{1}{x_2^2}\left(\frac{1}{4}-y^2\right)~.
\end{align}
The metric on $(x_1,x_2)$ becomes that of hyperbolic space, and a suitable quotient yields a compact Riemann surface of genus $g>1$.
At $y=0$, (\ref{eq:MN}) satisfies the boundary condition in (\ref{eq:bc}); at $y=\frac{1}{2}$ the function $e^D$ vanishes linearly and the $S^1$ shrinks smoothly, as in (\ref{eq:bc-yc}). The norm of the a-type solutions explicitly becomes
\begin{align}
 C_{\ell m n}^{\rm a}&=\frac{2^8\pi^2}{\kappa_{11}^2}\int dx_1 dx_2 dy |\varphi|^2 \frac{y^{2\ell}}{x_2^{2n+2}}\left(\frac{1}{4}-y^2\right)^n~.
\end{align}
For $n=0$ the $(x_1,x_2)$ integral becomes the volume of the Riemann surface, and can be evaluated in terms of the genus by the Gauss-Bonnet theorem.
The integrand is well behaved at $y=0$ for all $n$, but for $n<0$ it is not integrable at $y=\frac{1}{2}$, such that only the $n\geq 0$ modes are normalizable, in line with the general discussion below (\ref{eq:sol-2}).
Since holomorphic and harmonic functions defined everywhere on a compact Riemann surface are constant by the maximum principle, there is no degeneracy due to independent choices of $\varphi$.

\subsection{Bound on \texorpdfstring{$M^2$}{M}}\label{sec:M2-bound}

We now derive a general bound on $M^2$ for regular solutions that are normalizable with respect to (\ref{eq:C-M}).
To this end we start from (\ref{eq:spin-2-eq-M-4}), multiply by $\frac{1}{4}\bar\chi_{\ell m n} y^{2\ell+1}e^{(n+1)D}\partial_y D$ and integrate over $(x_1,x_2,y)$, which yields
\begin{align}
 \int d^2 x dy \bar\chi_{\ell m n}\Big[ &
 \partial_y y^{2+2\ell}e^{(n+1)D}\partial_y
 \nonumber\\&
 +y^{2\ell+2}(\partial_1-i \partial_2)e^{nD}(\partial_1+i\partial_2)
 -\frac{\mu^2}{4}y^{2\ell+1}e^{(n+1)D}(\partial_y D)
 \Big]\chi_{\ell m n}=0~.
\end{align}
We now integrate by parts in $y$ for the first term and in $x_1$, $x_2$ for the second term.
At $y=0$, $\partial_y D$ vanishes, and at $y=y_c$ the factors $e^{D}$ vanish.
Consequently, if $\phi$ is finite at $y=0$ and $y=y_c$ there are no boundary terms from the $y$ integration.
Likewise, if $(x_1,x_2)$ are coordinates on a closed surface, there are no boundary terms from the $(x_1,x_2)$ integration.
We thus find
\begin{align}
 \int d^2 x dy\, y^{2\ell}e^{n D} \Big[
 y^2 e^{D} \left|\partial_y\chi_{\ell m n}\right|^2
 +y^2\left|(\partial_1+i\partial_2)\chi_{\ell m n}\right|^2
 +\frac{\mu^2}{4}e^{D}(y\partial_y D) |\chi_{\ell m n}|^2
 \Big]&=0~.
\end{align}
For regular solutions $y\partial_y D$ is negative.\footnote{This is required for $f_4=-y^{-1}e^{2\tilde\lambda}\partial_y D$ to be positive, such that the spacetime signature is as desired.} Moreover, the last term in square brackets is proportional to the norm of $\phi_{\ell m n}$ (\ref{eq:C-M}), and in particular finite for normalizable solutions.
Since the first two terms in the square brackets are non-negative, $\mu^2$ is minimal in the case that they vanish, which leads to $\mu^2=0$. If $\chi_{\ell m n}$ is not constant, at least one of the first two terms is strictly positive, requiring $\mu^2>0$ for the equation to be satisfied. For regular solutions, we thus have $\mu^2>0$, or equivalently
\begin{align}\label{eq:M-bound}
 M^2&\geq  (2 \ell+n) (2 \ell+n+4)~.
\end{align}
In terms of the scaling dimension of the dual operator this bound becomes $\Delta\geq 4+2\ell+n$.

\section{Spin 2 fluctuations in Type IIA} \label{sec:IIA}

Assuming an additional isometry of the Toda equation (\ref{eq:toda-LLM}) in the $x_1,x_2$ plane, the M-theory solutions of sec.~\ref{sec:M-theory} can be reduced to obtain ten-dimensional Type IIA solutions. 
Since many explicit solutions  and their field theory  duals  are constructed directly in Type IIA, we will discuss the solution of the  PDE governing spin 2 fluctuations  in Type IIA explicitly.

\subsection{Type IIA supergravity solution} \label{sec:IIA-review}

Taking the additional isometry in the M-theory solution to be generated by $\partial_{x_1}$, the reduction is obtained by a change of coordinates.  As in \cite{Lin:2004nb} the function  $D$  and the coordinates $(x_2, y)$ are traded for a new function $V$ and   coordinates $(\sigma,\eta)$  as follows,
\begin{align}
 e^D&=\sigma^2~, & y &=\sigma\partial_\sigma V~, & x_2&=\partial_\eta V~.
\end{align}
Under this change of coordinates the Toda equation (\ref{eq:toda-LLM}) reduces to a linear PDE which is a cylindrical Laplace equation \cite{Ward:1990qt} for the function $V$,
\begin{align} \label{eq:V-eqn}
\ddot{V} + \sigma^2 V^{\prime\prime} = 0~, &
&\dot V&\equiv \sigma\partial_\sigma V~, &  V^\prime&\equiv \partial_\eta V~.
\end{align}
The string-frame metric and dilaton are given by
\begin{align} \label{metads}
ds^2 &= 4 f_1 ds^2_{AdS_5}+ f_2(d\sigma^2+d\eta^2) +f_3 ds_{S^2}^2 + f_4 d\beta^2~, & e^{2\phi} &= f_8~,
\end{align}
where
\begin{align}
 f_1 &= \left( {2 \dot V-\ddot V\over V''}\right)^{1\over 2}~, 
 &
 f_3 &=f_1 {2 V'' \dot V\over \tilde{\Delta}}~,
 &
 f_8 &= \left( { 4(2\dot V-\ddot V)^3\over V'' \dot V^2 \tilde{\Delta}^2}\right)^{1\over 2}~,
 \nonumber\\
 f_2 &= f_1{2V''\over \dot V}~, 
 & 
 f_4 &= f_1 {4V''\over 2 \dot V-\ddot V}\sigma^2~,
 &
 \tilde{\Delta} &=(2 \dot V-\ddot V) V''+(\dot V')^2~.
\end{align}

\subsection{Spin 2 fluctuations} \label{sec:spin-2-IIA}

We consider a traceless symmetric spin 2 fluctuation along the $AdS_5$ part of the Einstein frame metric, related to the string frame metric (\ref{metads}) by a factor $e^{-\phi/2}$,
\begin{align} \label{meta1}
ds^2 &= 4f_1e^{-\frac{\phi}{2}}\left(ds_{AdS_5}^2 + h_{\mu\nu}dx^\mu dx^\nu\right)+\hat g_{ab}dz^a dz^b~,
&
h_{\mu\nu} &= h^{[tt]}_{\mu\nu}(x) \psi(z)~,
\end{align}
where $\square^{(2)}_{AdS_5}h^{[tt]}_{\mu\nu} =(M^2-2) h^{[tt]}_{\mu\nu}$ and the metric on the internal space is
\begin{align}\label{eq:ghat-IIA}
\hat g_{ab}dz^a dz^b&= e^{-\phi/2}\Big( f_2(d\sigma^2+d\eta^2) +f_3 ds_{S^2}^2 + f_4 d\beta^2\Big)~.
\end{align}
The equation of motion for this perturbation $h$ in Einstein frame again takes the general form (\ref{eq:h-gen-eom}) \cite{Bachas:2011xa}.
With the decomposition into $h_{\mu\nu}^{[tt]}$ and $\psi$ as in (\ref{eq:h-eom}), eq.~(\ref{eq:h-gen-eom}) reduces to
\begin{align} \label{eq:IIA-1}
\frac{4}{(f_1e^{-\phi/2})^{3/2}\sqrt{\hat g}} \partial_a \left[(f_1e^{-\phi/2})^{5/2}\sqrt{\hat g} \hat g^{ab}  \partial_b\right] \psi=-M^2 \psi~.
\end{align}
With $\hat g$ in (\ref{eq:ghat-IIA}), this equation becomes
\begin{align} \label{eq:IIA-2}
\left[ {2\over V'' \dot V}   \tilde{\Delta} \nabla^2_{S^2} + \frac{2 \dot{V} - \ddot{V}}{V'' \sigma^2} \partial_\beta^2 + {2\over V'' \dot V \sigma} \partial_a \Big( \sigma \dot V^2 \partial^a\Big) + M^2 \right] \psi=0~.
\end{align}
The function $\psi$ is now expanded in spherical harmonics on $S^2$ and on the $S^1$ parametrized by $\beta$, and we introduce $\mu^2$ and define $\chi_{\ell m n}$ in analogy to the M-theory case
\begin{align} \label{eq:IIA-expand}
\psi &=\sum_{\ell m n}\phi_{\ell m n}Y_{\ell m} e^{in\beta}~,  
&
\phi_{\ell m n} &= \sigma^n\dot V^\ell \chi_{\ell m n}~, 
&
M^2&=\mu^2+(2 \ell+n) (2 \ell+n+4)~.
\end{align}
The real and imaginary parts again provide separate solutions.
Eq.~\eqref{eq:IIA-2} turns into
\begin{align} \label{eq:IIA-3} 
\frac{2}{\sigma^{2n-1}\dot V^{2\ell+1}\ddot V}\partial_a\left(\sigma^{2n+1} \dot V^{2+2\ell}\partial^a\chi_{\ell mn}\right)
 -\mu^2\chi_{\ell mn} &= 0~.
\end{align}
This equation is solved by $\chi_{\ell m n}=1$ with $\mu=0$ and for $n=0$ also by $\chi_{\ell m n}=V^\prime$ with $\mu=0$.
Paralleling the discussion in sec.~\ref{sec:M-theory}, this leads to the solutions
\begin{align}\label{eq:IIA-sols}
 \phi^{\rm a}_{\ell m n}&=\sigma^n\dot V^\ell \varphi(\sigma,\eta)~, & M^2 &= -4 + (2 + 2\ell +n)^2~,
 \nonumber\\
 \phi^{\rm b}_{\ell m n}&=\sigma^n\dot V^{-\ell-1}\varphi(\sigma,\eta)~, & M^2&=-4+(n-2\ell)^2~,
\end{align}
where $\varphi$ is constant for $n\neq 0$ and an arbitrary linear function in $V^\prime$, $\varphi=a+b V^\prime$, for $n=0$.
The latter correspond to the solutions with harmonic function $\varphi=a+bx_2$ in M-theory.
Moreover, there is a family of solutions for $\ell=0$ and non-negative $n\in\ZZ$,
\begin{align}\label{eq:phi-c-IIA}
\phi_n^{\rm c} &= \frac{\sigma^{n}}{\dot{V}}\left(\sigma^{-1}\partial_\sigma\right)^{n}\xi~, & M^2 &= -4+n^2~, & \ell &= 0~,
\end{align}
where $\xi$ is an arbitrary function satisfying $\ddot \xi + \sigma^2\xi^{\prime\prime}=0$. Examples for $\xi$ are $\ln\sigma$, $V$ and $\partial_\eta^p V$ for $p\geq 1$.
Since $M^2$ is below the bound (\ref{eq:M-bound}), these solutions are not regular.
For $\xi=\ln\sigma$, $\phi^{\rm c}_0$ corresponds to the M-theory solution in eq.~\eqref{eq:phi-c}.

\subsection{Regularity conditions} \label{sec:norm-IIA}
As in \cite{ReidEdwards:2010qs,Aharony:2012tz}, we take $0<\eta<\eta_c$ and $0<\sigma<\infty$. The boundary conditions for $V$ are
\begin{align}\label{eq:IIA-bc}
 \dot V\vert_{\eta=0,\eta_c}&=0~, & \dot V\vert_{\sigma=0}&=\lambda(\eta)~,
\end{align}
with continuous and piecewise-linear $\lambda$ encoding the details of the dual SCFT.
For solutions with multiple NS5-brane stacks the surface parametrized by $(\sigma,\eta)$ has  cuts along $\sigma$ at intermediate values of $\eta$ across which $V^\prime$ is discontinuous \cite{Aharony:2012tz}. On these cuts $\dot V$ vanishes.
Finiteness of the metric perturbation at $\eta=0$ and $\eta=\eta_c$ eliminates the $\phi_{\ell m n}^{\rm b}$ solutions.
The $\phi_{\ell m n}^{\rm a}$ are finite at $\eta=0$ and $\eta=\eta_c$, while finiteness at $\sigma=0$ imposes the constraint $n\geq 0$.
Since $V' \vert_{\sigma = 0}$ diverges when $\lambda' \neq 0$, $\varphi$ is restricted to be a constant for all $n$.
In summary, this leaves the $\phi^{\rm a}_{\ell m n}$ with $\ell,n\geq 0$ as regular solutions.

The norm is computed from the Einstein-frame action $S_{\rm IIA}=\frac{1}{2\kappa_{10}^2}\int \sqrt{-g}R+\dots$, in parallel to sec.~\ref{sec:norm-M}, by expanding to quadratic order in the perturbation. This yields
\begin{align}
\delta^2 S_{\text{IIA}} &= \sum_{\ell m n} C_{\ell m n} \int d^5 x \sqrt{-g_{AdS_5}} (h^{[tt]}_{\ell m n})^{\mu\nu} \left[ \Box^{(2)}_{AdS_5} + 2 - M^2 \right] (h^{[tt]}_{\ell m n})_{\mu\nu}~, 
\nonumber\\
C_{\ell m n } &= -\frac{2^7\pi^2}{\kappa_{10}^2}\int d\sigma d\eta\,(\partial_\sigma{\dot V}^2) (\phi_{\ell m n})^2~.
\end{align}
For the $\phi^{\rm a}_{\ell m n}$ solution with $\ell=m=n=0$, $C_{\ell m n}$ encodes the effective five-dimensional Newton constant and reduces to a boundary integral along $\eta$ in terms of $\lambda$ defined in (\ref{eq:IIA-bc}).

\section{Superconformal multiplets}\label{sec:multiplets}

Operators in unitary superconformal field theories are organized in unitary representations of the specific superconformal algebra. In this section we discuss how the spin 2 fluctuations found in the previous sections, and correspondingly the dual operators, fit into representations of the  $4d$ $\N{2}$ superconformal algebra $SU(2,2|2)$.

The representations of $SU(2,2|2)$ were worked out in \cite{Dolan:2002zh,Cordova:2016emh}, and we follow the conventions of \cite{Cordova:2016emh}. The  $SU(2,2|2)$ algebra contains an $SU(2)_R\times U(1)_R$ R-symmetry and we denote the $SU(2)_R$ Dynkin label by $R$ and the $U(1)_R$ charge by $r$. The Lorentz representation of the conformal algebra $SO(4,2)$ is labeled by two $SU(2)$ quantum numbers $j$ and $\bar j$ and the scaling dimension is denoted by $\Delta$. A state in a superconformal multiplet is then denoted by $[j,\bar j]_\Delta^{(R;r)}$. The  eight Poincare supersymmetry generators correspond to
\begin{align}\label{eq:Q}
Q:& \;  [1,0]^{1;-1}_{1\over 2}~, & \bar Q:& \;  [0,1]^{1;+1}_{1\over 2}~.
\end{align}
The representations of $SU(2,2|2)$ are combinations of two-sided  chiral multiplets of four types: $L, A_1,A_2$  and $B_1$. The $A_1,A_2$ and $B_1$ multiplets contain null states and obey strict bounds which fix the conformal dimension in terms of the other charges. A complete multiplet is then constructed by combining two chiral components,  for example   $A_2\bar A_2$ or $L\bar B_1$.

Holographically, the $SU(2)_R$ and $U(1)_R$ symmetries are realized as the isometries of $S^2$ and $S^1_\beta$, respectively.
The labels of the $S^2$ spherical harmonics, $\ell$, $m$, and the KK momentum, $n$,  are therefore related to the R-charges of the dual operator. The normalization is fixed by matching the charges of $Q$ in (\ref{eq:Q}) to those of the Killing spinors, which yields
\begin{align}
 R&=2\ell~, & r&=2n~.
\end{align}
With the scaling dimension identified via the standard holographic map, the complexified spin 2 fluctuations and their conjugates correspond to operators transforming as
\begin{align}\label{matchquant}
&[2,2]^{2\ell, \pm 2n}_{\Delta=4+2\ell+n}~, \qquad\qquad n \in \ZZ~, \qquad n\geq 0~.
\end{align}
The real metric fluctuations correspond to the real combinations.
Since supergravity does not contain higher-spin fields, the fluctuations around the solutions of \cite{Lin:2004nb}, and correspondingly the dual operators, fall into representations of $SU(2,2|2)$ with spin at most 2.
Using the tables in section 4.6 of \cite{Cordova:2016emh}, we can identify unique multiplets with spin at most $2$ containing the operators (\ref{matchquant}).
For $n=0$ the operators in (\ref{matchquant}) arise as $Q^2\bar Q^2$ descendants in $A_2\bar A_2$ multiplets with the scalar primary state
\begin{align}\label{a2a2}
A_2 \bar A_2:\quad [0,0]_{\Delta=2 + 2\ell}^{2\ell,0}~.
\end{align}
For $\ell=n=0$ the spin 2 operator in this multiplet is the stress tensor of the SCFT and the scalar primary has dimension $\Delta=2$.
For $n>0$ the operators in (\ref{matchquant}) arise as $Q^2\bar Q^2$ descendants in multiplets $A_2 \bar L$ and $L\bar A_2$ with scalar primaries
\begin{align}\label{a2lla2}
A_2 \bar L:& \quad [0,0]_{2 + 2\ell +n }^{2\ell,-2n}~, &  L\bar A_2:& \quad \quad [0,0]_{2 + 2\ell +n }^{2\ell,2n}~.
\end{align}
For $n=0$ the uniqueness of the multiplet (\ref{a2a2}) follows from the fact that 
any multiplet containing a $B_1$ (or $\bar B_1$) factor does not have a spin two operator, as the largest value of $j$ (or $\bar j $) is 1, while the multiplets $A_1 \bar A_1$ or $A_1\bar A_2$ contain higher spin states.
Similar arguments single out the choice (\ref{a2lla2}) for $n>0$.

Our results thus show that 4d $\N{2}$ SCFTs with holographic duals of the LLM type universally have operators in the multiplets (\ref{a2a2}) and  (\ref{a2lla2}). Identifying these operators in the field theory is an interesting question and we close with some comments.
Gauge theory descriptions of the field theories dual to the solutions of \cite{Lin:2004nb} typically involve long quivers with many $SU(N)$ gauge nodes \cite{Gaiotto:2009gz,Aharony:2012tz}.  
We will use a simpler $\N{2}$ gauge theory with a single $SU(N_c)$ gauge node and $N_F=2N_c$ fundamental hypermultiplets to illustrate the form of candidate scalar primary operators. This theory is not expected to have a simple holographic dual \cite{Gadde:2009dj,Gadde:2010zi} and we comment on the generalization to long quivers at the end.

We denote the hypermultiplet scalars by $q_{Ii}^a ,\bar q^{Ii}_{a}$, where $I$ is an $SU(2)_R$ index, $i=1,..,N_f$ a flavor index and $a$ a fundamental $SU(N_c)$ index.
The flavor contracted mesonic operators 
\begin{equation}\label{mdefa}
{\cal M}^{Ia}_{\; Jb} ={1\over \sqrt{2}} \sum_{i=1}^{N_f} q_{Ji}^a\bar q^{Ii}_{b}
\end{equation}
decompose into an $SU(2)_R$ singlet and a triplet,
\begin{align}
{\cal M}_{\bf 1} &= {\cal M}_I^I ~, & \big( {\cal M}_{\bf 3} \big)_I^J&= {\cal M}_I^J-{1\over 2}\delta_I^J {\cal M}_K^K~.
\end{align}
They further decompose into an adjoint and a singlet of $SU(N_c)$ and we use the adjoint in the following.
The scalar primary in the stress tensor multiplet is given by
\begin{equation}
{\cal T}= \phi \bar \phi - {\cal M}_1~,
\end{equation}
where $\phi$ is the complex scalar in the $SU(N_c)$ vector multiplet. $\phi$ is  an $SU(2)_R$ singlet, has engineering dimension $\Delta=1$ and, in the conventions of \cite{Cordova:2016emh}, $U(1)_R$ charge $2$.
From these fields we can construct a generalized single trace operator \cite{Gadde:2009dj}
\begin{equation}\label{worda}
\tr\Big({\cal T} ({\cal M}_3)^\ell \phi^n\Big)~,
\end{equation}
which  has  engineering dimension $\Delta= 2 + 2\ell+n$ and R-charge $(2\ell,2n)$. In  \cite{Gadde:2010zi}  it was shown that this operator picks up a one loop anomalous dimension and is not protected.  

For the long quiver theories with holographic duals of LLM type there are many more options to build similar operators. In addition to a large number of vector multiplets there are bifundamental hypermultiplets connecting adjacent gauge nodes and additional fundamental hypermultiplets. Those may be combined in various ways to form operators similar to (\ref{worda}). Our supergravity analysis suggests that among these operators there are protected ones corresponding to chiral primaries in the multiplets (\ref{a2a2}) and (\ref{a2lla2}) for all theories with holographic duals of LLM type. It would be interesting to find the exact form of these operators.

\clearpage

\section*{Acknowledgements}
We thank Oren Bergman and Thomas Dumitrescu for interesting discussions.
The work of M.G.  is supported in part by the National Science Foundation under grant PHY-16-19926. The work of C.F.U.  is supported by the Mani L. Bhaumik Institute for Theoretical Physics.

\bibliographystyle{JHEP}
\bibliography{4dn2}
\end{document}